\documentclass[sigconf]{acmart}

\AtBeginDocument{%
  \providecommand\BibTeX{{%
    \normalfont B\kern-0.5em{\scshape i\kern-0.25em b}\kern-0.8em\TeX}}}

\setcopyright{acmcopyright}
\copyrightyear{2018}
\acmYear{2018}
\acmDOI{XXXXXXX.XXXXXXX}

\acmConference[Submitted to SIGCSE '24]{Submitted to 2024 ACM SIGCSE Technical Symposium}{August 20--23, 2023}{Portland, OR, USA}
\acmPrice{15.00}
\acmISBN{978-1-4503-XXXX-X/18/06}

\newcommand{\courseyear}{2018}
\newcommand{\coursesemester}{Summer \courseyear}

\usepackage{tikz}
\usepackage{pgfplots}
\pgfplotsset{compat=1.13}

\usepackage{multirow}
\usepackage{pifont}
\usepackage{svg}

\usetikzlibrary{arrows.meta,automata,calc,fit,positioning,decorations.pathmorphing,decorations.text,decorations.pathreplacing,calligraphy,chains,shapes,matrix}
\tikzset{snake it/.style={decorate, decoration={snake,post length=0.5mm}}}
\tikzset{
	every state/.style={
		fill=gray!10,
		semithick
	},
	every edge/.style={
		draw,
		->,
		auto,
		very thick,
	},
	double distance=2pt,
	initial text={},
}

\begin{document}

\title{Designing Theory of Computing Backwards}

\author{Ryan E. Dougherty}
\email{ryan.dougherty@westpoint.edu}
\affiliation{%
  \institution{United States Military Academy}
  \city{West Point}
  \state{NY}
  \country{USA}
  \postcode{10996}
}

\renewcommand{\shortauthors}{Ryan E. Dougherty}

\begin{abstract}
The design of any technical Computer Science course must involve its context within the institution's CS program, but also incorporate any new material that is relevant and appropriately accessible to students.
In many institutions, theory of computing (ToC) courses within undergraduate CS programs are often placed near the end of the program, and have a very common structure of building off previous sections of the course. 
The central question behind any such course is ``What are the limits of computers?'' for various types of computational models.
However, what is often intuitive for students about what a ``computer'' is--a Turing machine--is taught at the end of the course, which necessitates motivation for earlier models.
This poster contains our experiences in designing a ToC course that teaches the material effectively ``backwards,'' with pedagogic motivation of instead asking the question ``What suitable restrictions can we place on computers to make their problems tractable?''
We also give recommendations for future course design.
\end{abstract}



\received{20 February 2007}
\received[revised]{12 March 2009}
\received[accepted]{5 June 2009}

\maketitle

\section{Introduction \& Motivation}

The over-arching question that is continually asked in nearly every Theoretical Computer Science (TCS) course is: ``what are the limitations of computers?'' 
The motivation is to start with ``simple'' machines, explore them, and to determine that they are not sufficiently powerful to handle problems that everyday computers can.
Then the course would add ``computational power'' to these machines so that they can more accurately compute answers to problems students should already know how to do with their own machines. 
Students then learn about decidability and undecidability, and observe that as computer models get more powerful, the fewer questions about them remain decidable, which is a trade-off. 
One of the main issues with this approach is the justification of \emph{which} ``simple'' machines to consider initially.

Therefore, we have developed such a course where we have re-thought the commonly taught ordering of topics. 
In particular, instead of asking the over-arching question located above, we asked a different question: ``what \emph{suitable restrictions} on computers can we place?''
Here, we start with the computational model that is equivalent to ``standard'' computers, and immediately find undecidability. 
Then we explore different restrictions to this model that (re)discover the less-powerful models. 
In the end, the conclusions and destination are the same--an understanding of computational model limitations--but the exploration is along a completely different road. 
Here we give our experience in designing a course in this way. 

\section{Course Context}\label{sec:course_context}

\begin{figure}
    \centering
    \includegraphics[width=0.9\columnwidth]{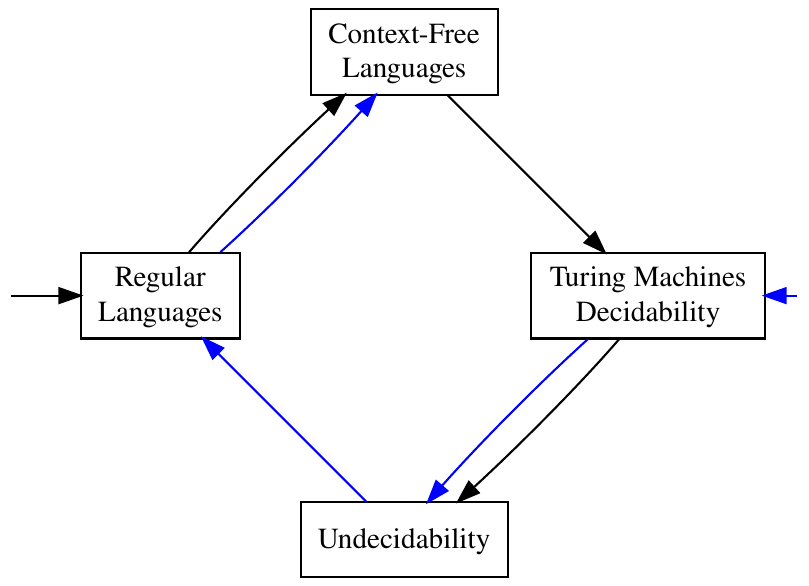}
    \caption{\label{fig:topic_ordering}Standard ToC course topic ordering (black arrows) and our proposed topic ordering (blue arrows).}
\end{figure}

The course normally involves four main topic sections, taught in the following order: (1) regular languages, (2) context-free languages, (3) Turing machines and decidability, and (4) undecidability. 
Each section involves constructing a machine or grammar for certain languages of each language type as well as proving that some languages do not fit into the appropriate type.
For the \coursesemester\ semester, we taught the course differently by instead approaching it from a different perspective.
Instead of building up more complex machines from simple ones, we started with complex machines that model everyday computers, and determined what appropriate restrictions one can place on the model that would make certain problems tractable. 
Specifically, we proposed a new order of the same four topics: $(3) \to (4) \to (1) \to (2)$\footnote{(4) here requires (3) for definitions.}. 
See Figure~\ref{fig:topic_ordering} for a visualization; the black arrows correspond to the standard ordering, and the blue arrows are our proposed ordering.
The arrows from nowhere indicates the topic where the associated course starts.

\section{Course Construction}\label{sec:course_construction}

This section contains our decisions on how we designed our ToC course, which was taught at Arizona State University during the Summer 2018 semester. 

Before this course we used the very popular ToC textbook by Sipser \cite{sipser2012introduction} as it is very standard, and also contains the concepts in the standard order.
However, at the time this course was offered, there did not exist textbooks that taught the concepts in our desired order.
Therefore, we needed to design our own lecture notes. 
We made an important decision in designing these notes in that many of the proofs and examples are intentionally left blank with an empty boxed area for students to fill in the proof/diagram.
Each of these blank sections is designed to flow naturally with the rest of the notes as if they were filled in.
Some theorems or examples that take less time to lecture upon are built-in to the notes and help provide students how to answer the theorem/example in a way we expect of them.

In the Sipser textbook, the formal definition for Turing Machines involves 7 parts: a set of states, two alphabets (input and tape), a transition function, a start state, and two halting states (for accepting and rejecting).
Even the finite automata definition with 5 parts historically was difficult for students to understand.
Instead of introducing the formal definition straight away (after motivation), we first do an example on how to build a TM.
This choice seems counter-intuitive, in that in mathematical courses the formal definition often comes first with examples to follow.
However, we have found that students can more easily derive formal definitions \emph{from} pictures than vice versa.
This section otherwise is straightforward, through the comparisons with Turing Machine variants (nondeterministic, multi-tape, etc.), to the Church-Turing thesis.

In the Sipser textbook, decidability and undecidability problems are phrased in terms of questions one can ask about a machine in a given model type.
Nearly every non-trivial question one can ask about Turing Machines is undecidable, partly due to Rice's theorem. 
However, Sipser's textbook contains all such questions grouped together in these sections.
An advantage in our approach is that these questions are spread out across the course sections.
In lessons we would create a table that contains check marks (for decidable) or ``X'' marks (for undecidable) for each model and associated question; for Turing Machines, all entries are X marks.
As further sections are encountered, more cells are filled in the table.
The table would be revisited so that students can make connections with past table entries. 
We believe it is important to have a compact visual diagram demonstrating the motivation for having different ``levels'' of models.

We then explain the important perspective they should have, namely in asking what kinds of \emph{restrictions} one can take to the model to make the problems tractable.
The Turing Machine model has several behaviors: (1) a read/write one-way infinite tape; (2) a tape head that moves left and right; and (3) a finite set of states with a transition function and some accept/reject criteria.
The last cannot be generalized to an infinite set of states nor removed entirely.
Therefore, some restriction needs to be made with the first two.
A single-tape machine whose tape is write-only is not interesting, and a tape head that only moves left is not interesting as no transitions can be taken without the machine crashing.

Therefore, there are four options: restrict the tape to be read-only, restrict the tape head to move right only, restrict the tape head to move left/right but only at the ``end'' of where the current work on the tape is, or restrict the tape to be a fixed and finite size (that must of course contain the input).
We explain to students that the first is difficult to prove but is nevertheless equivalent to a finite automaton. 
Students then learn about the machine model for the fourth--linear bounded automata (LBAs)--and the fundamental differences between them and TMs.
We then dive into the second option: only moving right on the tape.

Since our Turing Machine model only moves right, the notion of a tape is no longer needed; we then launch into standard terminology and proofs about regular languages and finite automata.
The difference is that throughout this the section we again introduce decidability and undecidability now in the context of regular languages. 
We quickly drop the visuals of a ``standard'' TM, both in the definition and examples, when defining finite automata as the extra computational power TMs have is unneeded. 

We then turn to context-free languages, which are equivalent to a TM that only has read/write operations performed at the ``end'' of where the current tape contents are. 
This section is also fairly standard, with again some decidability/undecidability results mixed in with common theorems and examples. 

\section{Future Work and Recommendations}\label{sec:future_work}
 
Does this topic ordering result in a statistically significant change in student course performance?
Our anecdotal experience indicates that there is not much of a difference, if any; ToC is a fundamentally hard course.
Just because the material is taught in a different way, all of the same hurdles need to be overcome. 

For practitioners who are looking at changes they want to make to their ToC course, we would strongly recommend to first determine its context within the CS curriculum and whether its needs are currently met.
Perhaps the topics themselves need to change; although easily well motivated, automata theory and computability are not the only topics within Theoretical Computer Science.
It may be time for the CSE community to reexamine ToC courses and whether they fit modern perspectives of theory and its applications. 

\begin{acks}
The opinions in the work are solely of the author, and do not necessarily reflect those of the U.S. Army, U.S. Army Research Labs, the U.S. Military Academy, or the Department of Defense.
\end{acks}

\bibliographystyle{ACM-Reference-Format}
\bibliography{main}

\end{document}